# Abelian Finite Group of DNA Genomic Sequences

Robersy Sanchez[1,2], Jesús Barreto[3], Eberto Morgado[4] and Ricardo Grau[2,4].

[1] Research Institute of Tropical Roots, Tuber Crops and Banana (INIVIT), Biotechnology group, Santo Domingo, Villa Clara, Cuba.
[2] Center of Studies on Informatics, Central University of Las Villas, Villa Clara, Cuba.
[3] Instituto Superior Pedagógico Félix Varela. Santa Clara. Villa Clara. Cuba.
[4] Faculty of Mathematics Physic and Computation, Central University of Las Villas, Villa Clara, Cuba

**Abstract**

The $Z_{64}$-algebra of the genetic code and DNA sequences of length $N$ was recently stated. In order to beat the limits of this structure −such as the impossibility of non-coding region analysis in genomes and the impossibility of the insertions and deletions analysis (indel mutations)− we have develop a cycle group structure over the of extended base triplets of DNA $X_1X_2X_3$, $X_i \in \{O, A, C, G, U\}$, where the letter O denote the base omission (deletion) in the codon. The obtained group is isomorphic to the abelian 5-group $Z_{125}$ of integer module 125. Next, it is defined the abelian finite group $S$ over a set of DNA alignment sequences of length $N$. The group $S$ could be represented as the direct sum of homocyclic groups: 2-group and 5-group. In particular, DNA subsequences without indel mutation could be considered building block of genes represented by homocyclic 2-groups (described in the previous $Z_{64}$-algebra). While those DNA subsequences affected by indel mutations are described by means of homocyclic 5-groups. This representation suggests identify genome block structures by way of a regular grammar capable of recognize it. In addition, this novel structure allows us a general analysis of the mutational pathways follow by genes and isofunctional genome regions by means of the automorphism group on $S$.

1. Introduction

The quantitative relationships between codons and between genes given in the molecular evolution process are an amazing challenge to mathematical biology. Regularities in the standard genetic code lead us to think this code has evolved in order to minimize the consequence of errors during transcription and translation (Epstein, 1966; Crick, 1968; Lewin, 2004). So, many attempts have been made to introduce a formal characterization of the genetic code (Jungck, 1978; Siemion et al., 1995; Jiménez-Montaño, 1999; Gillis, 2001). However, the most recent models lead us to go beyond the genetic code limits to deal with the quantitative relationship between DNA genomic sequences (Sanchez et al., 2004, 2005a, 2005c).

Recently was pointed out the $Z_{64}$-algebra of the genetic code ($C_g$) and the $N$-dimensional DNA sequences space $S$ defined on the set of all $64^N$ DNA sequences with $N$ codons (Sanchez et al., 2005c). The sum operation defined on this set is a manner to consecutively obtain all codons from the codon AAC (UUG) in such a way that the genetic code will represent a non-dimensional code scale of amino acids interaction energy in proteins. It was verified that most frequent mutation can be described by means of automorphisdms $f$: $(Z_{64})^N \rightarrow (Z_{64})^N$ over $S$. However, this model is limited to coding regions, while it is well known that in eukaryotes only a minute fraction of the genome –about 3%– called open reading frame (ORF) codes for proteins (Lewin, 2004). In addition, since non-coding DNA sequences can have a base pairs number not multiple of three, complete chromosomes and genomes can not be described by means of the $Z_{64}$-algebra $C_g$. Besides this, insertions and deletion mutations (indel mutations) in DNA sequences can not also be described by this algebra.

In order to beat these limitations, we propose an extension of the group ($C_g$, +) that we shall call extended group of the genetic code and next, we will show that all possible DNA sequence alignments with length *N* can be described by way of finite abelian groups which can decomposed into the direct sum of homocyclic 2-groups and 5-groups. A homocyclic group is a direct sum of cyclic groups of the same order. Any finite abelian group can be decomposed into a direct sum of homocyclic *p*-groups.

## 2. Extended genetic code group

A description of the genetic code abelian finite group ($C_g$, +) can be found in (Sanchez et al., 2005c). The extension of the codon set is achieved extending the source alphabet of the standard genetic code: {A, C, G, U}. To describe indel mutations in codons we have added the letter O to the set of the DNA bases, such that, for all extended triplet $X_1X_2X_3$, $X_i \in$ {O, A, C, G, U}. In a similar way to (Sanchez et al., 2005c) we can obtain the ordered set of extended triplets, presented in Table 1, and deduce a sum algorithm for extended triplets. The sum operation between two extended triplets *XYZ* and *X'Y'Z'* is obtained by means of base sum, presented in Table 2, which is also deduced from Table 1 (as in Sanchez et al., 2005c). The sum operation of extended triplets runs from the less biological significant base position –the third triplet position: *Z* and *Z´*– to the most important base position –located in the second position: *Y* and *Y´*– by means of the algorithm:

  i) Letters in the third positions are added according to the sum table (Table 2).
  ii) If the resultant letter of the sum operation is previous (in the order) to the added letters –the orders in the set of letters {O, A, C, G, U}–, then the new value is written and the base A is added to the next position.
  iii) The other letters are added according to the sum table, step 2, going from the first base to the second base.

We shall call the group defined in Table 2, the extended alphabet group of the four bases ($G_b$), and the group defined over the set of extended triplets, the extended genetic code group ($C_e$, +). Let us sum, for instance, the extended triplets OGC y UCG:

C + G = O, the third letters are added and the base A is added to the next position because letter O precedes bases C and G in the set of ordered extended bases {O, A, C, G, U}.
O + U + A = U + A = O, the first letters and the base A obtained in the first step are added. Again, base A is added to the next position.
G+C+A=O+A= A, the second bases are added to base A obtained in the second step. Finally, we have:

$$OGC + UCG = OAO$$

Since all finite cyclic groups with the same number of elements are isomorphic, then, group ($C_e$, +) is isomorphic to the group $Z_{125}$ of integers module 125, ($Z_{125}$, +). So, for instance, we can compute:

|   AGC   | ↔ |   82   |   AGC   | ↔ |   82   |   CCC   | ↔ |   62   |
|---|---|---|---|---|---|---|---|---|
| +UGU | ↔ | +99 | +AUA | ↔ | +106 | + AAU | ↔ | 34 |
|   ACA   | ↔ | 56 *mod* 125 |   CCG   | ↔ | 63 *mod* 125 |   UGA   | ↔ | 96 *mod* 125 |

Due to any abelian group is essentially a module over some ring, as in Sanchez et al. 2005c, we can say that the previous representation is the coordinate representation of the extended triplets from $Z_{125}$-Module ($C_e$, +) over the ring $Z_{125}$.

### 3. Abelian finite group of DNA sequences

Now, we can analyze the set of alignment sequences of length *N*. In the genomic DNA sequences are found open reading frames (ORF) that are building block of genes (Lewin, 2004). If we analyzed the multi-alignment of these sequences we can find subregions where there are not gaps introduced and we only found substitution mutations (see Fig. 1). These building blocks can correspond to complete exons or subregions, and can be described by means of the group ($C_g$, +). While, those genome regions where gaps appear –as a result of indel mutations– can be described by means of the monogen group ($C_e$, +). This is essentially an application of the fundamental theorem of abelian finite groups (Frobenius and Stickelberger, 1879; Dubreil and Dubreil-Jacotin,1963). By this theorem every finite abelian group *G* is isomorphic to a direct product of cyclic groups of prime power order. In particular, the theorem state a conical decomposition for every finite abelian group *G*, i.e. the group *G* is isomorphic to a direct product of cyclic groups

$$Z_{n_1} \times Z_{n_2} \times \cdots \times Z_{n_k}$$

such that $n_i \mid n_{i-1}$ for $i = 2, 3, \ldots, k$.

In the present case, as is showed in Fig 1, the group defined over the sequence space *S* –formed by the set of all $64^{m_1+m_2+\ldots+m_p}125^{n_1+n_2+\ldots+n_q}$ possible aligned sequences of length *N* ($N = n_1+\ldots+n_p+m_1+\ldots+m_q$)– is an heterocyclic group. This group split into a directed sum of homocyclic *p*-groups each one of them split into the direct sum of cyclic *p*-groups (monogens) with same order. Notice that for each fixed length *N* we can build manifold heterocyclic groups $S_i$, each one of them can have different decomposition into *p*-groups. So, we can characterize each group $S_i$ by means of their corresponding canonical decomposition into *p*-groups. That is, two sequences $S_1$ and $S_2$ could split into different homocyclic *p*-groups and, however, be isomorphic between them because have the same canonical decomposition. Biologically, such description is in correspondence with the fact that the new genetic information is created, simply, by way of reorganization of the genetic material in the chromosomes of living organisms (Lewin, 2004).

The automorphism group Aut(*S*) of the abelian group (*S*, +) is formed by set of invertible elements of the endomorphism ring End(*S*) of group (*S*, +). Since each finite abelian group *G* split into the direct sum of primary groups, the problem of the group structure of Aut(*G*) is reduced to the problem of the group structure Aut($G_p$), where $G_p$ denote the Sylow *p*-subgroups of *G*, for all prime numbers divisor of $|G|$.

**Tabla 1.** The set of ordered extended triplets. The bijection between the set of extended triplets and the set $Z_{125}$ is showed.

|   | No | O | No | A | No | C | No | G | No | U |   |
|---|----|---|----|---|----|---|----|---|----|---|---|
|   | 0  | OOO | 25 | OAO | 50 | OCO | 75 | OGO | 100 | OUO | O |
|   | 1  | OOA | 26 | OAA | 51 | OCA | 76 | OGA | 101 | OUA | A |
| O | 2  | OOC | 27 | OAC | 52 | OCC | 77 | OGC | 102 | OUC | C |
|   | 3  | OOG | 28 | OAG | 53 | OCG | 78 | OGG | 103 | OUG | G |
|   | 4  | OOU | 29 | OAU | 54 | OCU | 79 | OGU | 104 | OUU | U |
|   | 5  | AOO | 30 | AAO | 55 | ACO | 80 | AGO | 105 | AUO | O |
|   | 6  | AOA | 31 | AAA | 56 | ACA | 81 | AGA | 106 | AUA | A |
| A | 7  | AOC | 32 | AAC | 57 | ACC | 82 | AGC | 107 | AUC | C |
|   | 8  | AOG | 33 | AAG | 58 | ACG | 83 | AGG | 108 | AUG | G |
|   | 9  | AOU | 34 | AAU | 59 | ACU | 84 | AGU | 109 | AUU | U |
|   | 10 | COO | 35 | CAO | 60 | CCO | 85 | CGO | 110 | CUO | O |
|   | 11 | COA | 36 | CAA | 61 | CCA | 86 | CGA | 111 | CUA | A |
| C | 12 | COC | 37 | CAC | 62 | CCC | 87 | CGC | 112 | CUC | C |
|   | 13 | COG | 38 | CAG | 63 | CCG | 88 | CGG | 113 | CUG | G |
|   | 14 | COU | 39 | CAU | 64 | CCU | 89 | CGU | 114 | CUU | U |
|   | 15 | GOO | 40 | GAO | 65 | GCO | 90 | GGO | 115 | GUO | O |
|   | 16 | GOA | 41 | GAA | 66 | GCA | 91 | GGA | 116 | GUA | A |
| G | 17 | GOC | 42 | GAC | 67 | GCC | 92 | GGC | 117 | GUC | C |
|   | 18 | GOG | 43 | GAG | 68 | GCG | 93 | GGG | 118 | GUG | G |
|   | 19 | GOU | 44 | GAU | 69 | GCU | 94 | GGU | 119 | GUU | U |
|   | 20 | UOO | 45 | UAO | 70 | UCO | 95 | UGO | 120 | UUO | O |
|   | 21 | UOA | 46 | UAA | 71 | UCA | 96 | UGA | 121 | UUA | A |
| U | 22 | UOC | 47 | UAC | 72 | UCC | 97 | UGC | 122 | UUC | C |
|   | 23 | UOG | 48 | UAG | 73 | UCG | 98 | UGG | 123 | UUG | G |
|   | 24 | UOU | 49 | UAU | 74 | UCU | 99 | UGU | 124 | UUU | U |

**Table 2.** Sum table of the extended alphabet of four DNA bases {O, A, C, G, U}.

| + | O | A | C | G | U |
|---|---|---|---|---|---|
| O | O | A | C | G | U |
| A | A | C | G | U | O |
| C | C | G | U | O | A |
| G | G | U | O | A | C |
| U | U | O | A | C | G |

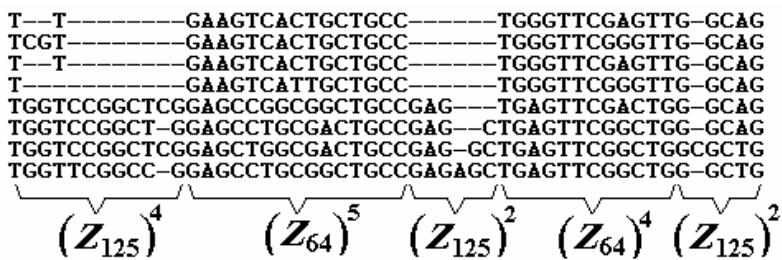

**Figura 1.** Building blocks from multiple sequence alignment of DNA sequences. Building block detection can achieved by means of the algorithm of multiple sequence alignments (Durbin et al., 1998; Baldi et al, 2001). Sequences, in this example, belong to an abelian group that split into the direct sum of 2-groups and 5-groups:

$$S = (Z_{5^3})^4 \oplus (Z_{2^6})^5 \oplus (Z_{5^3})^2 \oplus (Z_{2^6})^4 \oplus (Z_{5^3})^2$$

Following to Shoda (1928), Z.M. Kishkina showed that (Kurosch, 1955) if $G$ split into the direct sum $G = G_1 \oplus G_2 \oplus \ldots \oplus G_s$, then the endomorphism ring is isomorphic to the ring of all matrixes $(A_{ij})$, where $A_{ij} \in \text{Hom}(G_i, G_j)$, with usual sum and product of matrixes. In particular, the endomorphism that transform the DNA sequence $\alpha$ into $\beta$ ($\alpha, \beta \in S$) is represented by a matrix whose elements in the principal diagonal are matrixes $A_{ii} \in \text{End}(G_i)$ (or $A_{ii} \in \text{Aut}(G_i)$) and out of the principal diagonal are null matrixes. In biological terms, mutations in genomic sequences will correspond to automorphism when $A_{ii} \in \text{Aut}(G_i)$, for all $i$ in the representing matrix, and this will happen when $\det(A_{ii}) \not\equiv 0 \bmod p_i$ (since $G_i$ is a $p_i$-group (Shoda, 1928)). As was pointed out in (Sanchez et al., 2005c) this fact will be possible when all mutant DNA subsequences $\beta_i$ keep the order of the corresponding sequences of the will type $\alpha_i$ ($\alpha_i, \beta_i \in G_i$). A natural gene always satisfies this algebraic condition.

The last representation as a direct sum of powers of $Z_{64}$ and $Z_{125}$ suggests that we would associate to building block structure of genomes a regular grammar. This grammar would recognize the expression of the form:

$$S = (Z_{5^3})^{n_1} \oplus (Z_{2^6})^{m_1} \oplus (Z_{5^3})^{n_2} \oplus (Z_{2^6})^{m_2} \oplus ... \oplus (Z_{5^3})^{n_p} \oplus (Z_{2^6})^{m_p}$$

with $n_1$ and $m_p$ more than or equal to cero, and $n_i$ and $m_i$ strictly positives. It would be enough consider a regular grammar like this:

$S \to x\, X;\ S \to y\, Y;\ X \to x\, X;\ Y \to y\, Y;\ X \to y\, Y;\ Y \to x\, X;\ X \to \lambda;\ Y \to \lambda$

Where $S$ is the start state, $X$ and $Y$ are terminal state, which are acceptation states in the corresponding finite automata ($\lambda$ is the chain end), "$x$" and "$y$" represent, in this case, "$x=Z_{125}$" y "$y=Z_{64}$". It is well know that the regular grammar have important applications in bioinformatics in sequence analyzes (Baldi et al., 2001 and Durbin et al., 1998). This analysis will not exclude from this approach.

### 4. Conclusions

Limitations of the $Z_{64}$-Module algebraic structure of DNA sequences lead us to define the extended triplet set of the genetic code $C_e$ using an extension of the four letter alphabet of DNA molecule {O, A, C, G, U}. In the extended triplet set was defined the group $(C_e, +)$ isomorphic to the cyclic group $Z_{125}$. As a result, given an iso-functional genomic DNA region, the sequence alignment of the set of all iso-functional genomic DNA sequences, can be represented as a finite group. This group split into the direct sum of 2-groups and 5-groups:

$$S = (Z_{5^3})^{n_1} \oplus (Z_{2^6})^{m_1} \oplus (Z_{5^3})^{n_2} \oplus (Z_{2^6})^{m_2} \oplus ... \oplus (Z_{5^3})^{n_p} \oplus (Z_{2^6})^{m_p}$$

Such decomposition allows us characterize the more frequent mutational pathway follow by DNA sequences in the set of all $64^{m_1+m_2+...+m_p} 125^{n_1+n_2+...+n_p}$ possible aligned sequences of length $N$. In particular, mutational pathway can be represented by means of automorphism where the elements of the representing matrix $A$ in the principal diagonal are matrixes $A_{ii} \in \text{Aut}(G_i)$ and out of the principal diagonal are null matrixes.

In addition, since for any length we can find manifold heterocyclic groups $S_i$, the last representation suggests the possible recognition of block structures in genomes by means of a regular grammar.

**References**

Baldi, Pierre; Brunak, Soren; Bioinformatics, the Machine Learning Approach, 2nd. Edition, MIT Press, Cambridge, England, pp 277-297 (2001)

Crick, F.H.C.: The origin of the genetic code. J. Mol. Biol. 38, 367-379 (1968).

Dubreil, P., Dubreil-Jacotin, M. L. Lecciones de álgebra moderna. Editorial Reverté (1963).

Durbin, Richard; Eddy, Sean R.; Krogh, Anders; Mitchison, Graeme; Biological sequence analysis. Probabilistic models of proteins and nucleic acids, Cambridge University Press, , pp. 233-259 (1998).

Epstein, C. J.: Role of the amino-acid "code" and of selection for conformation in the evolution of proteins. Nature 210, 25-28 (1966).

Freeland, S.J., Knight, R.D., Landweber, L.F., Hurst, L.D. Early fixation of an optimal genetic code. Mol. Biol. Evol. 17, 511-8 (2000).

Frobenius F G; Stickelberger: Ueber Gruppen von vertauschbaren Elementen. Journal für die reine und angewandte Mathematik. 86. S.217-262 Z 655 – 86 (1879).

Gillis, D., Massar, S., Cerf, N.J., Rooman, M.: Optimality of the genetic code with respect to protein stability and amino acid frequencies. Genome Biology 2, research0049.1–research0049.12 (2001).

Jiménez-Montaño, M.A.,. Protein Evolution Drives the Evolution of the Genetic Code and Vice Versa. BioSystems 54: 47-64 (1999).

Jungck, J.R., "The genetic code as a periodic tables", J.Mol.Evol. 11, 211-224 (1978).

Karasev, V.A., Stefanov, V.E.: Topological Nature of the Genetic Code. J. Theor. Biol. 209, 303-317 (2001).

Kurosch, A. The theory of groups, 2nd ed., Chelsea Publ. Co., New York, 1955.

Lewin, B. Genes VIII. Oxford University Press. (2004).

Sánchez, R., Morgado, E., Grau, R.: The Genetic Code Boolean Lattice. MATCH Commun. Math. Comput. Chem 52, 29-46 (2004)

Sánchez, R., Morgado, E., Grau, R.: A Genetic Code Boolean Structure. I. The Meaning of Boolean Deductions. Bull. Math. Biol. 67, 1–14 (2005a).

Sánchez, R., Perfetti, L. A., R. Grau y E. R. Morgado. A New DNA Sequences Vector Space on a Genetic Code Galois Field. MATCH Commun. Math. Comput. Chem. Vol. 54, 1, 3-28 (2005b).

Sánchez, R., Morgado, E., Grau, R. Gene algebra from a genetic code algebraic structure. J. Math. Biol. 51, 431 - 457, (2005c).

Shoda K.: Über die Automorphismen Einer Endlichen Abelichen Gruppe. Math. Ann. 100, 674-686 (1928).

Siemion, I.Z., Siemion, P.J., Krajewski, K.: Chou-Fasman conformational amino acid parameters and the genetic code. Biosystems 36, 231-238 (1995).